\documentclass[aps,prd,superscriptaddress,showkeys,showpacs,twocolumn,a4paper]{revtex4-1}
\usepackage{ulem,amssymb}
\usepackage{cancel}
\usepackage{color}
\usepackage{graphicx}
\usepackage{epsfig}
\usepackage{mathrsfs}
\usepackage{amsmath,amsthm}
\usepackage[dvipsnames]{xcolor}
\usepackage{float}
\usepackage[bottom]{footmisc}
\definecolor{darkblue}{RGB}{0,0,196}
\definecolor{darkgreen}{RGB}{0,120,0}
\usepackage[colorlinks=true,linktocpage=true,linkcolor=darkblue,citecolor=red,urlcolor=darkblue]{hyperref}

\newcommand{\be}{\begin{equation}}
\newcommand{\ee}{\end{equation}}
\newcommand{\ba}{\begin{eqnarray}}
\newcommand{\ea}{\end{eqnarray}}

\begin{document}
\title{Thermal and thermoelectric responses of hot QCD medium in time-varying magnetic fields}
\author{Gowthama K K}
\email{k$_$gowthama@iitgn.ac.in}
\affiliation{Indian Institute of Technology Gandhinagar, Gandhinagar-382355, Gujarat, India}

\author{Manu Kurian}
\email{manu.kurian@mail.mcgill.ca}
\affiliation{Indian Institute of Technology Gandhinagar, Gandhinagar-382355, Gujarat, India}
\affiliation{Department of Physics, McGill University, 3600 University Street, Montreal, QC, H3A 2T8, Canada}

\author{Vinod Chandra}
\email{vchandra@iitgn.ac.in}
\affiliation{Indian Institute of Technology Gandhinagar, Gandhinagar-382355, Gujarat, India}

\begin{abstract}
The thermal response of the hot QCD matter has been studied in the presence of a time-varying magnetic field. The impact of magnetic field, its time dependence, and the collision aspects of the medium on thermal transport have been studied within the relativistic kinetic theory. The decay time of the magnetic field in the medium seems to have a strong dependence on thermal conductivity. The applicability of the Wiedemann-Franz law for the QCD medium has been investigated in the presence of time-varying external electromagnetic fields. The phenomenological significance of thermal transport in heavy-ion collision experiments has also been investigated by relating the thermal conductivity to the elliptic flow through the Knudsen number. The investigations are extended to study the thermoelectric behavior of hot QCD medium and its dependence on the magnetic field. The time dependent magnetic field is observed to significantly influence the thermoelectric behavior of the medium.
\end{abstract}
\maketitle

 \section{Introduction}
The prime focus of experiments at Relativistic Heavy Ion Collider (RHIC) and Large Hadron Collider (LHC) is to study and characterise the properties of strongly interacting matter: the Quark-Gluon Plasma (QGP)~\cite{Adams:2005dq, Back:2004je,Arsene:2004fa,Adcox:2004mh,Aamodt:2010pb}. The evolution of the QGP produced in these experiments is modeled by relativistic hydrodynamics, and transport coefficients of the medium act as the input parameters governing the evolution~\cite{Gale:2013da, Heinz:2013th}. The theoretical description of the hadron elliptic flow within dissipative hydrodynamics at collision experiments at RHIC and LHC provide insights into various transport processes in the QCD medium. Advances are made on the phenomenological constraints on shear and bulk viscosities of the QGP medium and is an interesting area of contemporary physics~\cite{Teaney:2003kp,Romatschke:2007mq,Ryu:2015vwa}. 

Recent measurements on directed flow of $D^0/\bar{D}^0$ at the RHIC and LHC gave indications of the presence of a strong magnetic field in the early stages of collision experiments~\cite{Acharya:2019ijj,Adam:2019wnk}. The strength of the generated magnetic field is expected to be in the order of $10^{18}-10^{19}$ G~\cite{Skokov:2009qp,Zhong:2014cda,Voronyuk:2011jd,Deng:2012pc} and decay in space and time, depending on the electromagnetic responses of the medium. However, a proper framework to model the evolution of the generated field in the QGP medium is not completely developed so far. Several recent studies inferred that the magnetic field might persist for a longer time in the medium and will affect the equilibrium and non-equilibrium properties of the medium~\cite{Tuchin:2013apa,McLerran:2013hla, Stewart:2021mjz,Tuchin:2019gkg,Yan:2021zjc}. QCD thermodynamics and momentum transport processes have been studied both in the weak magnetic field~\cite{Dash:2020vxk,Das:2019pqd, Kurian:2020qjr, Rath:2022oum,Ghosh:2021knc} and strong field regimes~\cite{Karmakar:2019tdp, Hattori:2017qih,Kurian:2018qwb,Dey:2021fbo,Ghosh:2022xtv, Ghosh:2018cxb}. Several attempts have been made to explore the electromagnetic responses of the medium to constant and inhomogeneous fields~\cite{Feng:2017tsh,Thakur:2019bnf,Dey:2019axu,Hattori:2016lqx,Fukushima:2017lvb, Dash:2020vxk,Kurian:2017yxj,Kurian:2019fty,Ghosh:2019ubc, Gowthama:2020ghl}. 

Recent investigations~\cite{Liu:2020dxg, Fu:2022myl} suggested the significance of non-vanishing temperature gradient over the spatial separations of non-equilibrium fluid and chemical potential gradient in the QCD medium in the context of spin Hall current at the heavy-ion collision experiments. The current study focuses on the thermal and thermoelectric transport processes due to the temperature gradient in the QCD medium in the presence of a time-evolving magnetic field. 

The thermal transport process has received less attention as compared to the momentum and electric charge transport in the QCD medium. This is due to the fact that the baryon number is not significant in the high energetic collision experiments. But, the baryon number and chemical potential can not be neglected for the lower energetic collision experiments at RHIC beam energy scan and for upcoming programs at the Facility for Antiproton and Ion Research (FAIR). The thermal response of the medium due to the temperature gradient has been studied~\cite{Denicol:2012vq,Kapusta:2012zb,Kadam:2017iaz,Khan:2020rdw}, and the heat current is shown to be along the direction of the local temperature gradient. The phenomenological significance of thermal transport on the elliptic flow at collision experiments has been studied~\cite{Bhalerao:2005mm}. The inclusion of a constant magnetic field generates an extra `Hall-like' component along the direction perpendicular to both the magnetic and the thermal driving force~\cite{Das:2019pqd,Pushpa:2021pgb}. The current study is on the thermal transport process of the QCD medium with a background time-varying magnetic field. The focus of the regime is where the time inhomogeneity of the external magnetic field is not large such that the collisional aspects in the medium cannot be neglected in the analysis. A general decomposition of heat current in the presence of a time-varying magnetic field is formulated. The relative significance of thermal response to the electric charge transport in the QCD medium is analyzed. The validity of the Wiedemann-Franz law is studied for the chosen range of temperature. The dependence of the magnetic field evolution in the medium on thermal conductivity and its phenomenological significance are also explored in the analysis.

Thermoelectric response of the QCD medium in the presence of a time-varying magnetic field is another aspect of focus. The decaying magnetic field will give rise to an induced electric field and hence electric current in the QCD medium. In addition to that, the temperature gradient in an electrically conducting medium generates the electric current, which can be described as the thermoelectric or Seebeck effect. The thermoelectric properties of the condensed matter systems have been well explored and recently started receiving attention for the QCD medium~\cite{Bhatt:2018ncr}. In Refs.~\cite{Das:2020beh, Zhang:2020efz, Dey:2021crc, Kurian:2021zyb}, the impact of the magnetic field on Seebeck and Nernst coefficients has been investigated. In the current study, a general formalism to describe the thermoelectric behavior of the QCD medium is presented while considering the time evolution of the magnetic field.   

The manuscript is organized as follows. In Section \ref{II}, the formalism for the thermal transport of the QCD medium in a time-varying weak magnetic field is described within the transport theory. Section \ref{III} is devoted to discussions on the phenomenological significance of thermal transport in heavy-ion collision experiments. In Section \ref{IV}, the thermoelectric behavior of the QCD medium is discussed in the presence of a time-evolving magnetic field. The results and the followed discussions on the thermal and thermoelectric responses of the QCD medium are presented in Section \ref{V}. Finally, the analysis is concluded with an outlook in Section \ref{VI}.

{{\it{ Notations and conventions}:}  The subscript $k$ represents the particle species. The electric charge of $k$-th species particle with flavour $f$ is denoted by $q_{f_k}$. The gluonic degeneracy factor $g_g =\sum_f 2N_c$ and $g_{q/\bar{q}} = 2N_c N_f$ for quarks and anti quarks, with $N_f$ being the number of flavours. The velocity of the particles is defined as ${\bf v}=\frac{{\bf p}}{\epsilon_k}$, where ${\bf p}$ is the momentum and $\epsilon_k$ is the energy. The vector components are denoted by the Latin indices $A^i$ for the vector ${\bf A}$. The quantity $E = |{\bf E}|$ and $B = |{\bf B}|$ denote the magnitude of the electric and magnetic fields, respectively.}
\section{Thermal transport in hot QCD medium}\label{II}
The energy-momentum tensor $T^{\mu\nu}$ and particle flow $N^{\mu}$ of the QGP medium can be defined in terms of quarks/antiquarks and gluonic momentum distribution function as follows,
\begin{align}\label{1.1}
T^{\mu\nu}(x)=\sum_{k}g_k\int{d{P}_k\,{p}_k^{\mu}\,{p}_k^{\nu}\,f_k(x,{p}_{k})},
\end{align}
and
\begin{align}\label{1.2}
N^{\mu}(x)=\sum_{k}g_k\int{d{P}_k\,{p}_k^{\mu}\,f_k(x,{p}_{k})},
\end{align}
respectively. Here, $d{P}_k\equiv\frac{d^3\mid{\bf{{p}}}_k\mid}{(2\pi)^3\epsilon_{k}}$ is the integral measure. The near-equilibrium momentum distribution function of the medium particles can be defined as,
\begin{align}\label{1.3}
&f_k = f^{0}_k +\delta f_k, &&f^0_k=\frac{1}{1\pm\exp{\big(\beta( \epsilon_k \mp \mu)\big)}},
\end{align}
where $f^{0}_k$ is the equilibrium part and $\delta f_k$ measures the non-equilibrium correction to the momentum distribution with $\delta f_k/f^0_k \ll1$. Employing Eq.~(\ref{1.3}) in Eq.~(\ref{1.1}) and Eq.~(\ref{1.2}),  the macroscopic quantities can be represented with equilibrium and non-equilibrium parts as $T^{\mu\nu}=T^{0~\mu\nu}+\Delta T^{\mu\nu} (x)$ and $N^{\mu}=N^{0~\mu}+\Delta N^{\mu} (x)$. The thermal response of the hot QCD medium can be studied in terms of dissipative net heat flow. The heat current for single component particle can be defined as, 
\begin{align}\label{1.4}
{ I^i_k}= \Delta T^{0i}_k -h_k \Delta N^i_k,  
 \end{align}
where $h_k$ is the enthalpy per particle. 
Employing Eq.~(\ref{1.1}) and Eq.~(\ref{1.2}), the microscopic definition of heat flow takes the form as,
\begin{align}\label{1.5}
   {\bf I}_k = \sum_k \int d{P}_k {\bf p}_k (\epsilon_k -h_k)\delta f_k.
\end{align}
The non-equilibrium part of the distribution function can be obtained by solving the Boltzmann equation by choosing the appropriate collision integral and has the following form, 
\begin{align}\label{1.51}
{p}^{\mu}_k\,\partial_{\mu}f_k(x,{p}_k)+\Big(q_{f_k}F^{\mu\nu}{p}_{k\,  \mu}\Big)\partial^{(p)}_{\mu} f_k=C[f_k],
\end{align}
where $C[f_k]$ is the collision kernel of the form, $C[f_k] =-\frac{\delta f_k}{\tau_{R_k}}$ in the relaxation time approximation (RTA) and $F^{\mu\nu}$ is the electromagnetic field strength tensor. We now proceed to discuss the various cases of the magnetic field.\\

{\it{Case 1,}} ({\bf B}=0): Within the  RTA, an iterative Chapman-Enskog like solution can be obtained for $\delta f_k$ as~\cite{Kurian:2020qjr},
\begin{align}\label{1.6}
\!\!\delta f_k &= \tau_{R_k}\Bigg[\beta \bigg\{{p}_k^0~\partial_0 T + {p}_k^i\,\partial_i T \bigg\} \!+\! \frac{T}{{p}^0_k} \!\bigg\{{p}^0_k\,\partial_0\Big(\frac{\mu}{T}\Big) \nonumber\\
&\!+\! {p}_k^i\,\partial_i \Big(\frac{\mu}{T}\Big) \bigg\}\!-\! \frac{1}{{p}^0_{k}} \!\bigg\{{p}^0_{k}\, {p}_{k}^\nu\,\partial_0 u_\nu \!+\! {p}^i_{k}\, {p}_{k}^\nu\,\partial_i u_\nu \bigg\}\! \Bigg]\frac{\partial f^0_k}{\partial\epsilon_k},
\end{align}
where $\tau_{R_k}$ is the thermal relaxation time. The $\delta f_k$ can be decomposed into different independent thermodynamic forces such as bulk pressure, shear viscous, and thermal driving forces. The current focus is on the thermal driving force due to the temperature gradient in the medium, $X_{i}=\frac{\partial_i T}{T}-\frac{\partial_i P}{n_k h_k}$. The Eq.~(\ref{1.6}) can be further simplified by employing relativistic Gibbs-Duhem relation, $\partial_i \Big(\frac{\mu}{T}\Big)=-\frac{h_k}{T^2}(\partial_iT-\frac{T}{n_k h_k}\partial_iP)$, and energy-momentum conservation equation as,
\begin{align}\label{1.8}
\!\!\delta f_k = \tau_{R_k}\frac{\partial f^0_k}{\partial\epsilon_k}\big(\epsilon_{k}-h_k\big){\bf v}_k.{\bf X} + \delta f_{k\,{\text{shear}}}+\delta f_{k\,{\text{bulk}}}.
\end{align}

{\it{Case 2,}} ({\it constant} {\bf B}): In a weakly magnetized QGP, the magnetic field can be considered as a perturbation in the system as its strength is much lesser than that of the temperature scale of the medium. By solving the Boltzmann equation in the presence of a weak magnetic field within RTA, the non-equilibrium part of the distribution due to the thermal driving force can be defined as follows~\cite{Kurian:2020qjr},
\begin{align}\label{1.9}
\delta f_k=&\tau_{R_k}\frac{\big({\epsilon_{k}-h_k}\big)}{(1+\tau_{R_k}^2\, \Omega^2_{c\, k})}\bigg[\big({\bf v}_k.{\bf X}\big)+\tau_{R_k}\, \Omega_{c\, k} {\bf v}_k. \big({\bf X}\times {\bf b}\big)\nonumber\\
&+\tau_{R_k}\, \Omega^2_{c\, k}\,\big({\bf b}.{\bf X}\big)\,\big({\bf v}_k.{\bf b}\big)\bigg]\,\frac{\partial f^o_k}{\partial\epsilon_k},
\end{align}
where $\Omega_{c\, k}=\frac{q_{f_k} \mid{\bf B}\mid}{\epsilon_k}$ and ${\bf b}$ is the direction of the magnetic field. It is important to emphasize that for a strongly magnetized medium, the charged particle will have $1+1$-dimensional constrained Landau level dynamics, which is beyond the scope of the current analysis.\\

{\it{Case 3,}} ({\it time evolving} {\bf B}): The evolution of the magnetic field affects the transport process in the conducting medium. We start with an ansatz for $\delta f_k$ in the presence of time dependent magnetic field as,
\begin{equation}\label{1.10}
\delta f_k=({\bf{p}}.{\bf \Xi} ) \frac{\partial f^0_k}{\partial \epsilon_k}, 
\end{equation} 
where ${\bf{\Xi}}$ can be defined in terms of the external perturbation and its time derivative that can deviate system slightly away from equilibrium. We consider the leading order source terms as $\textbf{B}, \textbf{X}, (\textbf{X}\times \textbf{B}), \dot{\textbf{B}}, (\textbf{X}\times \dot{\textbf{B}})$. As we considered the case of a slowly varying magnetic field, the terms with two and higher order space-time derivatives are neglected.
Hence, the quantity ${\bf{\Xi}}$ takes the form as follows,
\begin{align}\label{5.9}
\mathbf{\Xi} =& \alpha_1\textbf{B}+ \alpha_2\textbf{X}+ \alpha_3(\textbf{X}\times \textbf{B})+ +\alpha_4 \dot{\textbf{B}} +\alpha_5(\textbf{X}\times \dot{\textbf{B}}).
\end{align} 
The unknown functions $\alpha_{i}$ ($i=(1, 2,.., 5)$) can be obtained by substituting Eq.~(\ref{5.9}) in the Boltzmann equation. Employing Eq.~(\ref{5.9}) in Eq.~(\ref{1.51}), we obtain
%
\begin{align}\label{10}
& \epsilon_k {\bf v}.\Big[\dot{\alpha_1 }{\bf B}+\alpha_1 \dot{{\bf B}}+\dot{\alpha_2} {\bf X}+\alpha_3 ({\bf X}\times \dot{{\bf B}}) +\dot{\alpha_3 }({\bf X} \times {\bf B})+\dot{\alpha_4} \dot{\textbf{B}}
\nonumber\\
&+\alpha_4 \ddot{\textbf{B}}+\dot{\alpha_5 }({\bf X} \times \dot{{\bf B}}) +\alpha_5 (\textbf{X} +\ddot{\textbf{B}})\Big]-(\epsilon_k -h_k)\textbf{v}.\textbf{X}\nonumber\\
&-\alpha_2 q_{f_k}\textbf{v}.(\textbf{X} \times \textbf{B}) +\alpha_3 q_{f_k}\textbf{v}.\textbf{X}(\textbf{B}.\textbf{B})-\alpha_3 q_{f_k}\textbf{v}.\textbf{B}(\textbf{B}.\textbf{X})\nonumber\\
&+\alpha_5q_{f_k}\textbf{v}.\textbf{B}(\dot{\textbf{B}}.\textbf{X}) -\alpha_5 q_{f_k}\textbf{v}.\dot{\textbf{B}}(\textbf{B}.\textbf{X})=-\frac{\epsilon_k}{\tau_R}\Big[\alpha_1 {\bf v}.{\bf B}\nonumber\\
&+\alpha_2 {\bf v}.{\bf X}+\alpha_3 {\bf v}.({\bf X}\times {\bf B})+\alpha_4 {\bf v}.\dot{\textbf{B}}+\alpha_5 {\bf v}.({\bf X}\times \dot{{\bf B}})\Big].
\end{align}
%
Comparing the tensorial structures on both sides of Eq.~(\ref{10}) and solving the followed coupled equations, we can obtain $\alpha_i$. The detailed calculation of $\alpha_i$ is presented in Appendix~\ref{AppendixA}.  
The  general forms of the coefficients $\alpha_i$ are obtained as follows,  
\begin{align}\label{11}
 &\alpha_1 =-\frac{i(\textbf{B}.\textbf{X})}{F \epsilon_k} I_0 e^{\eta_0}+\frac{i(\textbf{B}.\textbf{X})}{2F \epsilon_k} I_1 e^{\eta_1} +\frac{i(\textbf{B}.\textbf{X})}{2F \epsilon_k} I_2 e^{\eta_2},\\
 &\alpha_2 =-i\frac{F}{2 \epsilon_k} I_1 e^{\eta_1}+i\frac{F}{2 \epsilon_k} I_2 e^{\eta_1} ,\label{20.2}\\
 &\alpha_3 =\frac{ I_1}{2 \epsilon_k} e^{\eta_1} +\frac{I_2}{2 \epsilon_k} e^{\eta_1},\label{20.4}\\
 &\alpha_4 = -\tau_R \alpha_1 -\frac{q_{fk} \tau_R^2}{\epsilon_k} \alpha_3 (\textbf{B}.\textbf{X}),\label{20.1}\\
 &\alpha_5 =-\tau_R \alpha_3\label{11.11},
\end{align}
where the functions $\eta_j$ and $I_j$ as defined in Eq.~(\ref{15}), depend on the profile of the magnetic field evolution.  
Further, employing Eq.~(\ref{1.10}) in Eq.~(\ref{1.5}), the heat current in the presence of time decaying magnetic field can be defined as, 
\begin{align}\label{}
\textbf{I} = \kappa_0 T \textbf{X} +\kappa_1 T (\textbf{X} \times \textbf{B}) + \kappa_2 T (\textbf{X} \times \dot{\textbf{B}}),   
\end{align}
with $\kappa_0$, $\kappa_1$, and $\kappa_2$ as thermal transport coefficients.
It is important to emphasize that the coefficients $\alpha_1, \alpha_4$ vanish due to the parity considerations and with the choice of direction of the magnetic field in the medium.
The components of the heat current will dependent on the magnetic field evolution in the medium. To that end, we have considered two different profiles for the time dependent magnetic field.
\begin{figure*}
    \centering
    \centering
    \hspace{-2.5cm}
    \vspace{0.5cm}
    \includegraphics[width=0.485\textwidth]{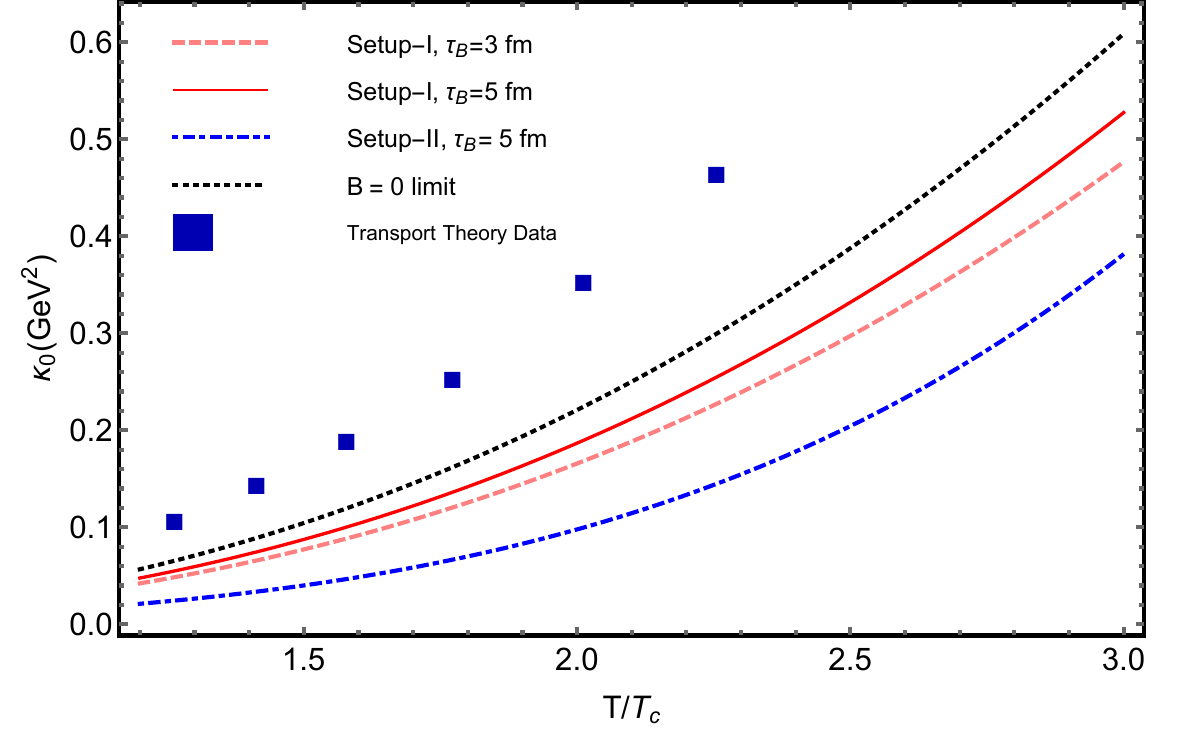}
    \vspace{-1.cm}
    \includegraphics[width=0.585\textwidth]{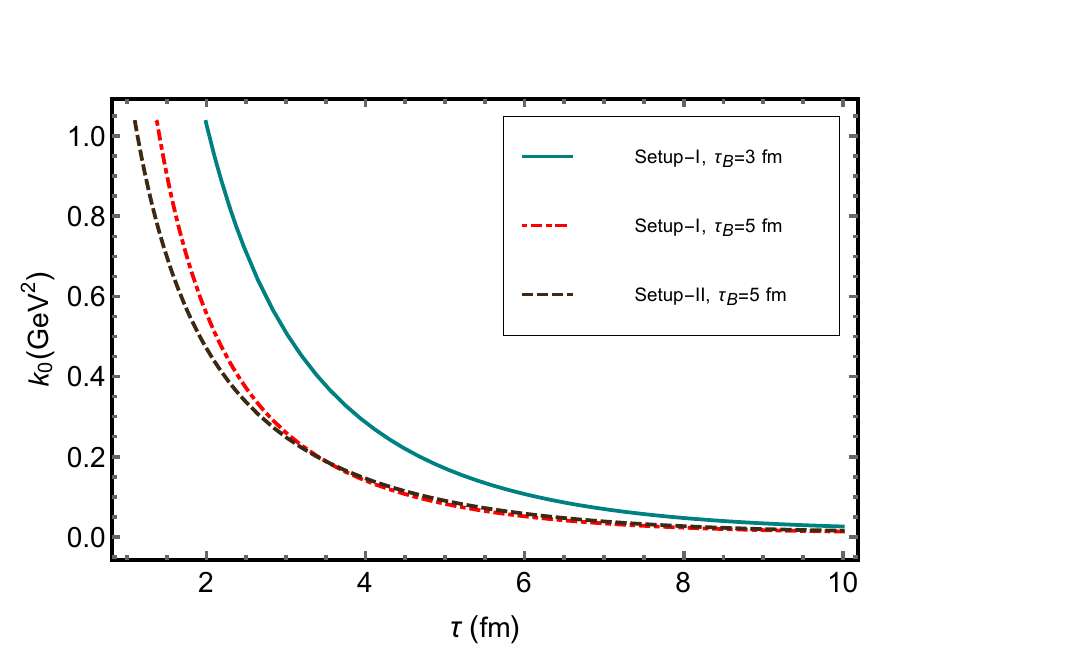}
    \hspace{-2.5cm}
    \caption{{(Left panel) Temperature dependence of thermal conductivity $\kappa_0$ for two different choices of magnetic field. The result is compared with the transport theory estimation with constant magnetic field~\cite{Rath:2021ryd}. (Right panel) The proper time evolution of $\kappa_0$ in the QGP medium. We consider $eB_0=0.08$ GeV$^2$ and $\mu=100$ MeV for the analysis.  }}
\label{f1}
\end{figure*}
\subsubsection*{Setup-I}
Here, the form of the magnetic field is adopted from Refs.~\cite{Hongo:2013cqa, Satow:2014lia} as,
\begin{align}\label{II.19}
 eB =  eB_0 \rho(\tau)\rho_B(r),
 \end{align}
where $\rho(\tau)=e^{-\frac{\tau}{\tau_B}}$ with $\tau_B$ as the decay parameter of the magnetic field, describes the evolution of the field with respect to the proper time. {A higher value of $\tau_B$ indicates a slowly varying magnetic field, and a lower value represents a rapidly decaying field. In the presence of a conducting medium, the magnetic field is expected to vary slowly, $i.e.$, with a higher value of $\tau_B$.} As the focus is on the time-evolving fields, the spatial distribution $\rho_B(r)$ of the field is neglected in the present analysis. We consider the case with inhomogeneity of the external field to be small ($i.e$, a larger value of decay parameter $\tau_B$) such that collisional aspects of the QCD medium can not be neglected. In this limit, the cyclotron frequency (proportional to the strength of the magnetic field $eB$ is in the same range of inverse of decay parameter $\tau_B$. This condition imposes constraints on the strength of the magnetic field (to be in the weak field limit) and inhomogeneity of the field (to be in the case of a magnetic field with small inhomogeneity). Further, we fix the direction of the temperature gradient and magnetic field in the medium for the quantitative estimation. For the case where magnetic field direction is transverse to the temperature gradient in the medium and in the limit where cyclotron frequency $\Omega_k$ is approximately equal to the inverse of the magnetic decay time, we obtain,
\begin{align}\label{12}
    &\eta_j = -\frac{\tau}{\tau_R} +a_j\Bigg(i\frac{\sqrt{1+\frac{\tau_R}{\tau_B}}}{\tau_B}\tau \Bigg),
\end{align}
\begin{align}
    &I_j = \frac{1}{\sqrt{1+\frac{\tau_R}{\tau_B}} B_0} \frac{e^{\Big(\frac{1}{\tau_R} +\frac{1}{\tau_B}-a_ji\frac{\sqrt{1+\frac{\tau_R}{\tau_B}}}{\tau_B} \Big)\tau}}{\Big( \frac{1}{\tau_R} +\frac{1}{\tau_B}-a_ji\frac{\sqrt{1+\frac{\tau_R}{\tau_B}}}{\tau_B} \Big)}.\label{12.1}
\end{align}
The coefficients $\alpha_i$ for the particular choice of magnetic field are obtained by substituting Eq.~(\ref{12}) and Eq.~(\ref{12.1}) in Eqs.~(\ref{11})-(\ref{1.111}). Employing the forms of distribution function and magnetic field as in Eq.~(\ref{II.19}), the heat current takes the form as, 
\begin{align}\label{1.111}
\textbf{I} = \kappa_0 T \textbf{X} +\big(\bar{\kappa_1}+ \bar{\kappa_2}\big) T  (\textbf{X} \times \textbf{b}),  
\end{align}
where $\textbf{b}$ is the direction of the chosen $\textbf{B}$ and $\dot{\textbf{B}}$. 
The first term in Eq.~(\ref{1.111}) denotes the leading order contribution to heat current in the medium. The thermal conductivity $\kappa_0$ takes the form as follows,
\begin{align}\label{18}
    &\kappa_0 =  \frac{1}{3T} \sum_k g_k \int dP_k\frac{ p^2_k}{\epsilon_k}(\epsilon_k -h_k)^2 \frac{M_1}{M_1^2 +M_2^2}(-\frac{\partial f_k^0}{\partial \epsilon_k}), 
\end{align}
where the effects of magnetic field evolution are entering through $M_1 =\big(\frac{1}{\tau_R} +\frac{1}{\tau_B}\big)$, $M_2=\frac{\sqrt{1+\tau_R/\tau_B}}{\tau_B}$. The presence of the magnetic field in the medium modifies $\kappa_0$ and, in the limit of ${\bf B}=0$ and constant magnetic field, the thermal conductivity $\kappa_0$ reduce to the forms in Ref.~\cite{PhysRevD.96.094003} and Ref.~\cite{Das:2019pqd}, respectively. The coefficients $\bar{\kappa}_1$ and $\bar{\kappa}_2$ associated with the thermal transport arises due to the magnetic field in the medium, and take the following forms,
\begin{align}\label{}
    &\bar{\kappa}_1 = \frac{1}{3T} \sum_k g_k \int dP_k\frac{ p^2_k}{\epsilon_k}(\epsilon_k -h_k)^2 \frac{1}{\tau_B (M_1^2 +M_2^2)}(-\frac{\partial f_k^0}{\partial \epsilon_k}),\label{23.1} \\
     &\bar{\kappa}_2 = \frac{1}{3T} \sum_k g_k \int dP_k\frac{ p^2_k}{\epsilon_k}(\epsilon_k -h_k)^2 \frac{\tau_R}{\tau_B^2(M_1^2 +M_2^2)}(-\frac{\partial f_k^0}{\partial \epsilon_k}).\label{}
\end{align}
In Eq.~(\ref{23.1}), the term with $\bar{\kappa}_1$ describes the `Hall-like' thermal response. The magnetic field can induce anisotropy in the transport processes in the medium. Similar to the case of electric charge transport, in the presence of a magnetic field, there will be a flow of particles in the direction perpendicular to the magnetic field and source of perturbation (temperature gradient in the case of thermal transport, and electric field in the case of electric charge transport) due to the Lorentz force term. This corresponds to a Hall-like heat current in the thermal transport process. The component $\bar{\kappa}_2$ is the additional component that arises due to the chosen time decay of the magnetic field in the medium. We have employed the decomposition of heat current with $\kappa_0$, $\bar{\kappa}_1$, and $\bar{\kappa}_2$ to highlight each contributions separately. In the limit of constant magnetic field ($\tau_B\rightarrow \infty$), $\bar{\kappa}_2$ vanishes. Also, in the case of a vanishing magnetic field, the term associated with both $\bar{\kappa}_1$ and $\bar{\kappa}_2$ vanishes and reduces back to the isotropic result. Further, we proceed to explore the dependence of the choice of magnetic evolution on thermal transport in the medium.
\subsubsection*{Setup-II}
Here, we have adopted the form of the magnetic field as in Ref.~\cite{Sun:2020wkg} in which the time dependence of the field can be described as follows,
\begin{align}
   B(\tau)=\frac{eB_0}{1+\tau/\tau_B}.
\end{align} 
Following the similar formalism as employed in the case of setup-I, the functions $\alpha_i$ can be obtained for this particular choice of the magnetic field evolution. The heat current takes the same form as described in Eq.~(\ref{1.111}) for the setup-II with the form of conductivities as follows,
\begin{align}\label{}
    &\kappa_0 =  \frac{1}{3T} \sum_k g_k \int dP_k\frac{ p^2_k}{\epsilon_k}(\epsilon_k -h_k)^2\Bigg[1+ \frac{A_1}{A_2} \Bigg](-\frac{\partial f_k^0}{\partial \epsilon_k}),\label{26.1}\\ 
    &\bar{\kappa}_1 = \frac{1}{3T} \sum_k g_k \int dP_k\frac{ p^2_k}{\epsilon_k}(\epsilon_k -h_k)^2 \frac{(1-1/\tau_B)}{A_2}(-\frac{\partial f_k^0}{\partial \epsilon_k}),\label{} \\
     &\bar{\kappa}_2 = \frac{1}{3T} \sum_k g_k \int dP_k\frac{ p^2_k}{\epsilon_k}(\epsilon_k -h_k)^2 \frac{\tau_R(1-1/\tau_B)}{\tau_BA_2}(-\frac{\partial f_k^0}{\partial \epsilon_k}),\label{}
\end{align}
where $A_j (j=1, 2)$ takes the form as $A_1 =\frac{\epsilon_k}{q_{fk} \tau_B}\big(\frac{\tau_B}{\tau_R} -\frac{1}{\tau_R} -\frac{\tau_B}{\tau_R^2}-\frac{\sqrt{1+\frac{\tau_R \epsilon_k}{q_{fk} B_0 \tau_B^2}}}{\tau_B}\big)$, $A_2= \tau_B \big( \frac{1}{\tau_R^2} +\frac{\sqrt{1+\frac{\tau_R \epsilon_k}{q_{fk} B_0 \tau_B^2}}}{\tau_B^2}\big)$.
\begin{figure*}
    \hspace{-3cm}
    \includegraphics[width=0.485\textwidth, height =0.3\textwidth]{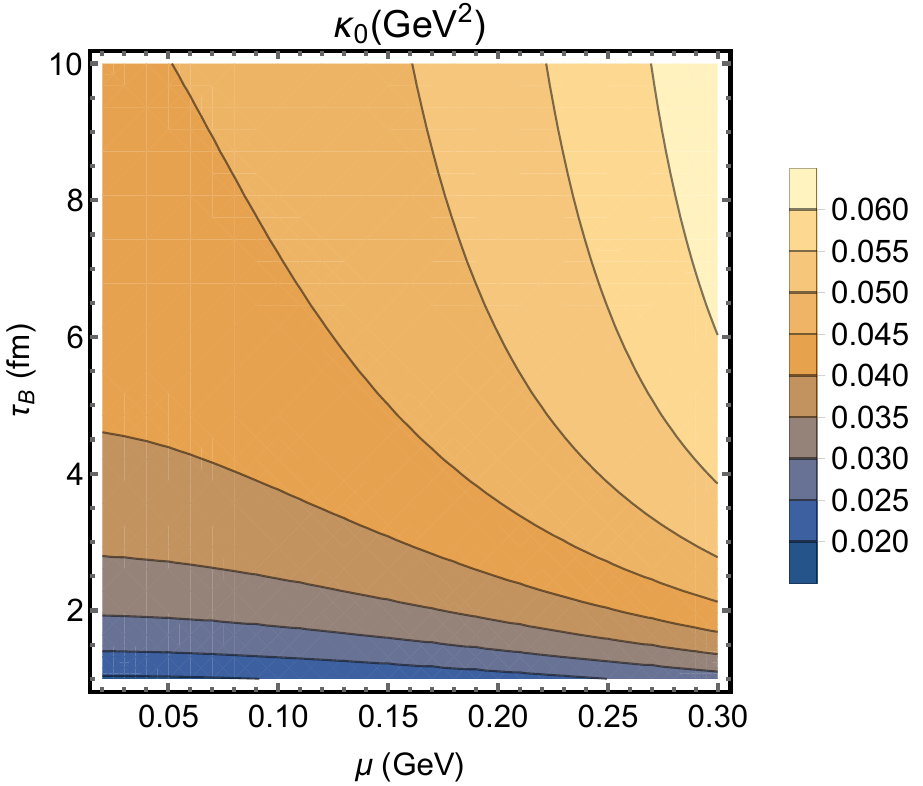}
    \includegraphics[width=0.48\textwidth]{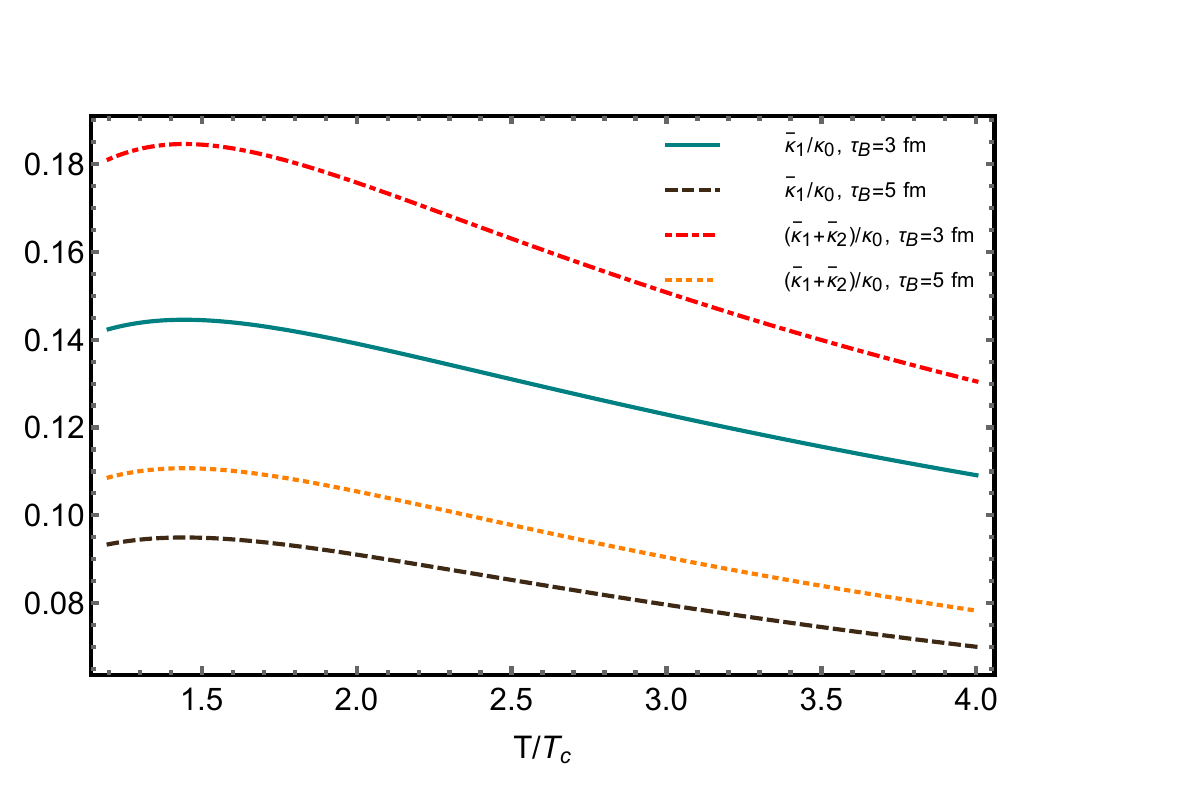}
    \hspace{-3.5cm}
    \caption{(Left panel) The effect of chemical potential and magnetic field decay parameter on $\kappa_0$ at a constant temperature $T=200$ MeV. Curved lines denote constant value contours of $\kappa_0$. (Right panel) Temperature dependence of the ratio of `Hall-like' thermal conductivity $\bar{\kappa}_1$ and $\bar{\kappa}_1+\bar{\kappa}_2$ to $\kappa_0$, for two different choices of magnetic field at $\mu=100$ MeV.}
\label{f2}
\end{figure*}
\section{Phenomenological significance of thermal transport in heavy-ion collision experiments}\label{III}
In this section, we consider the significance of thermal transport in the presence of a time-evolving magnetic field in the context of collision experiments. The thermal conductivity obtained can be employed to study the Knudsen number~\cite{Rath:2021ryd}. The impact of thermal conductivity of the medium on the elliptic flow is explored in Ref.~\cite{Bhalerao:2005mm} for the case of a vanishing magnetic field. The current study focus on the dependence of magnetic field evolution on thermal transport and its impact on the elliptic flow coefficient. Another aspect is the comparative study of thermal and electric charge transport in the QCD medium. The relative significance of thermal conductivity to electrical conductivity can be quantified in terms of Lorenz number $L$.
\subsection{Knudsen number and elliptic flow}
Knudsen number, $Kn$, is defined as the ratio of the mean path ($\lambda$) of the constituent particle to the size of the system, $l$,
\begin{align}\label{}
    Kn=\frac{\lambda}{l}.
\end{align}
If $Kn$ is equal to or greater than one, then the mean free path is comparable to the system size, and the continuum assumption of hydrodynamics is no longer applicable. Hydrodynamical modeling applies when the mean free path is less than the system size, $Kn << 1$. The mean free path is related to the thermal conductivity as $\lambda = \frac{3\kappa_0}{vC_v}$
where $v$ is the relative speed, and $C_v$ is the specific heat at constant volume. Hence, the Knudsen number can be expressed in terms of thermal conductivity as, 
\begin{align}\label{21}
    Kn = \frac{3\kappa_0}{lvC_v}.
\end{align}
For the quantitative estimation, we have chosen $v \approx 1$ and $l = 1$ fm~\cite{Rath:2022oum}.
The elliptic flow $v_2$ can be expressed in terms of the Knudsen number as~\cite{Bhalerao:2005mm}, 
\begin{align}\label{22}
    v_2 = \frac{v_2^h}{1+\frac{Kn}{Kn_0}},
\end{align}
where $v_2^h$ is the elliptic flow at the hydrodynamical limit, $Kn \rightarrow 0$, the quantity $Kn_0$ is a number obtained to fit the Monte-Carlo simulations of the relativistic Boltzmann equation~\cite{Bhalerao:2005mm}. In our present analysis we have taken $v_2^h =0.3 \pm 0.02$ and $Kn_0 = 0.7$~\cite{Gombeaud:2007ub}. The effects of a time-varying magnetic field on the thermal conductivity and hence the elliptic flow are discussed in detail in the results and discussion section.

\subsection{Thermal versus electric charge transport}
The relative significance of thermal and charge transport can be studied through the Wiedemann-Franz law, characterized by the Lorenz number $L$,
\begin{align}
    L = \frac{\text{Thermal conductivity}}{\text{Electrical conductivity} \times T}.
\end{align}
The value of $L$ indicates if the medium is a good thermal or electrical conductor. In condensed matter physics, it is observed that for metals, the Lorenz number, $L$, is a constant as the temperature varies. This section aims to study the behavior of the Lorenz number as a function of temperature in the presence of a time-varying, weak magnetic field and to explore the validity of the Wiedemann-Franz law in different directions. 

To study the charge transport of the medium in the presence of an external electric field, $\textbf{E}$ is introduced in the direction transverse to that of the magnetic field. The electric current density in the QCD medium can be defined as~\cite{K:2021sct}, 
\begin{align}\label{23}
   \textbf{j} = j_e \hat{\textbf{e}} +j_H (\hat{\textbf{e}} \times \hat{\textbf{b}}). 
\end{align}
The  Ohmic current $j_e$ is along the direction of the electric field $\hat{\textbf{e}}$, and  Hall current $j_H$ is perpendicular to both the electric and magnetic field. For the case of a time-evolving magnetic field, we have $j_e=j_e^{(0)}$ and $j_H=j_H^{(0)}+j_H^{(1)}$ with
\begin{align}\label{25}
    &j_e^{(0)} =  \frac{2E}{3} N_c \sum_k \sum_f (q_{fk})^2 \int dP_k\frac{ p^2_k}{\epsilon_k^2}(-\frac{\partial f_k^0}{\partial \epsilon_k}) {N}_2,\\ 
     &j_H^{(0)} = \frac{2E}{3} N_c \sum_k \sum_f (q_{fk})^3\int dP_k \frac{ p^2_k}{\epsilon_k^3}(-\frac{\partial f_k^0}{\partial \epsilon_k}) N_1,\label{31.2}\\ 
     &j_H^{(1)} = \frac{2E}{3\tau_B} N_c \sum_k \sum_f (q_{fk})^3\int dP_k \frac{ p^2_k}{\epsilon_k^3}(-\frac{\partial f_k^0}{\partial \epsilon_k}) \tau_R N_1,\label{31.4}
\end{align}
where $N_j (j=1, 2)$ functions can be defined as $N_1 =\big(\frac{1}{\tau_R} +\frac{1}{\tau_B}\big) N$, $N_2=-\big(\tau_R N_1-\frac{\tau_R^2}{\tau_B^2}N\big)/\big(1+(\frac{\tau_R}{\tau_B})^2\big)$ with, 
\begin{align}\label{26}
 N = \Bigg[{ \frac{1}{ \tau_R} +\frac{1}{\tau_B}+\frac{\sqrt{1+\frac{\tau_R}{\tau_B}}}{\tau_B} }\Bigg]^{-1}.
\end{align}
The electric conductivities can be written in terms of the currents as, $\sigma_e = j_e^{(0)}/E,$ $\sigma_H^{(0)} = j_H^{(0)}/EB$ and, $\sigma_H^{(1)} = j_H^{(1)}/EB $. It is important to note that if we switch on the time dependence of the external electric field, the current density will have more components with ${\bf\dot{E}}$ and ${\bf\dot{E}}\times {\bf\dot{B}}$ as source terms. It can be seen that the thermal and electric charge transport processes are analogous to each other. The thermal driving force is the source of the response in thermal transport, whereas the electric field plays a similar role in the case of charge transport. The coefficient $\kappa_0$ is analogous to the Ohmic conductivity associated with the electric charge transport in the medium. Notably, similar to the behavior of $\bar{\kappa}_1$ and $\bar{\kappa}_2$, Hall conductivity and its correction due to time-varying magnetic field tends to zero in the limit of vanishing chemical potential. The temperature and time-varying magnetic field dependence of the Lorentz number are discussed in the results section. Now, we proceed with the thermoelectric behavior of the QCD medium in the presence of a time-evolving magnetic field.

\begin{figure}
    \hspace{-2.5cm}
    \includegraphics[width=0.58\textwidth]{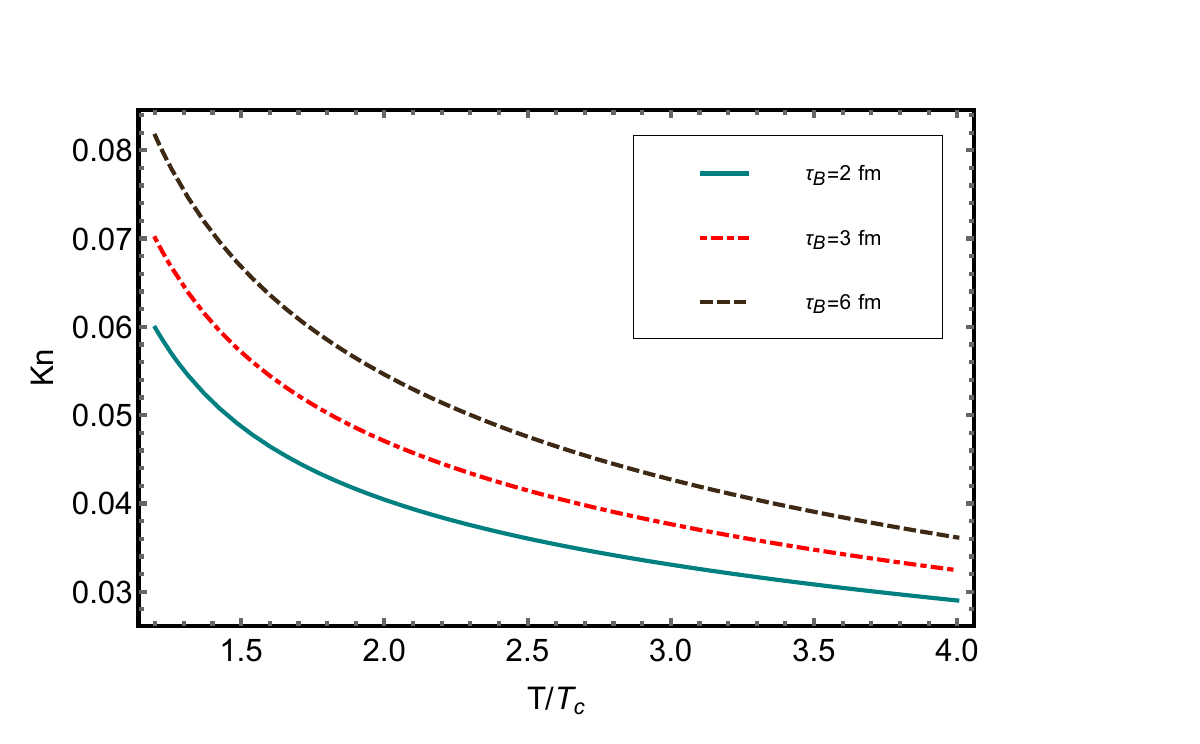}
    \hspace{-3cm}
    \caption{The temperature behavior of Knudsen number $Kn$ for various values of magnetic field decay parameter, $\tau_B= 2,3,6$ fm at $\mu =100$ MeV.}
\label{f3}
\end{figure}
\begin{figure}
    \centering
    \centering
    \hspace{-3cm}
    \vspace{-1.5cm}
    \includegraphics[width=0.52\textwidth]{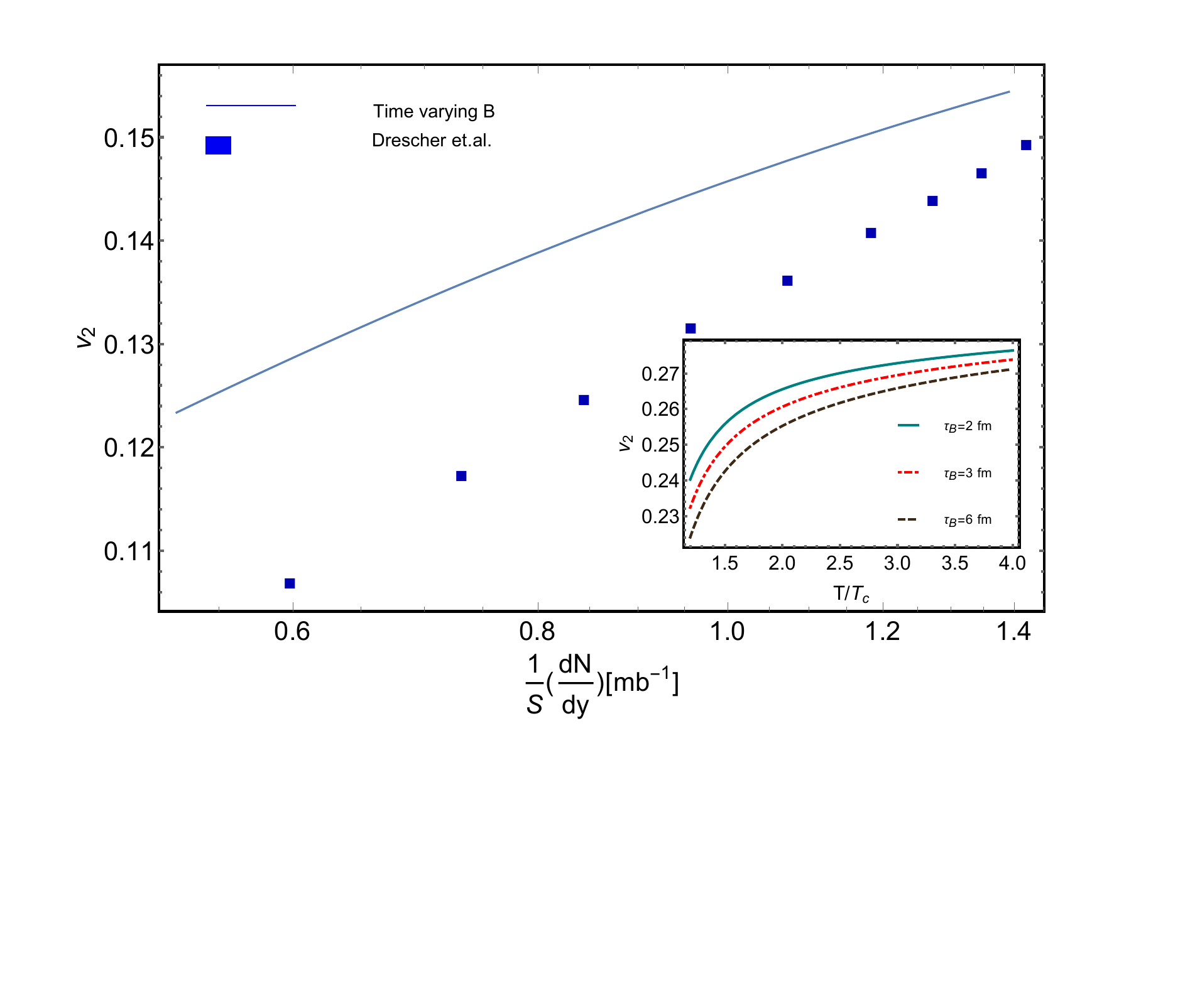}
    \hspace{-2.5cm}
    \vspace{-.5 cm}
    \caption{Elliptic flow $v_2$ as a function of multiplicity. The result is compared with that from ~\cite{Drescher:2007cd}. Inset plot: The temperature dependence of $v_2$ for various magnetic field decay parameter values, $\tau_B= 2,3,6$ fm at $\mu =100$ MeV.  }
\label{f4}
\end{figure}
\begin{figure*}
    \centering
    \includegraphics[width=0.45\textwidth]{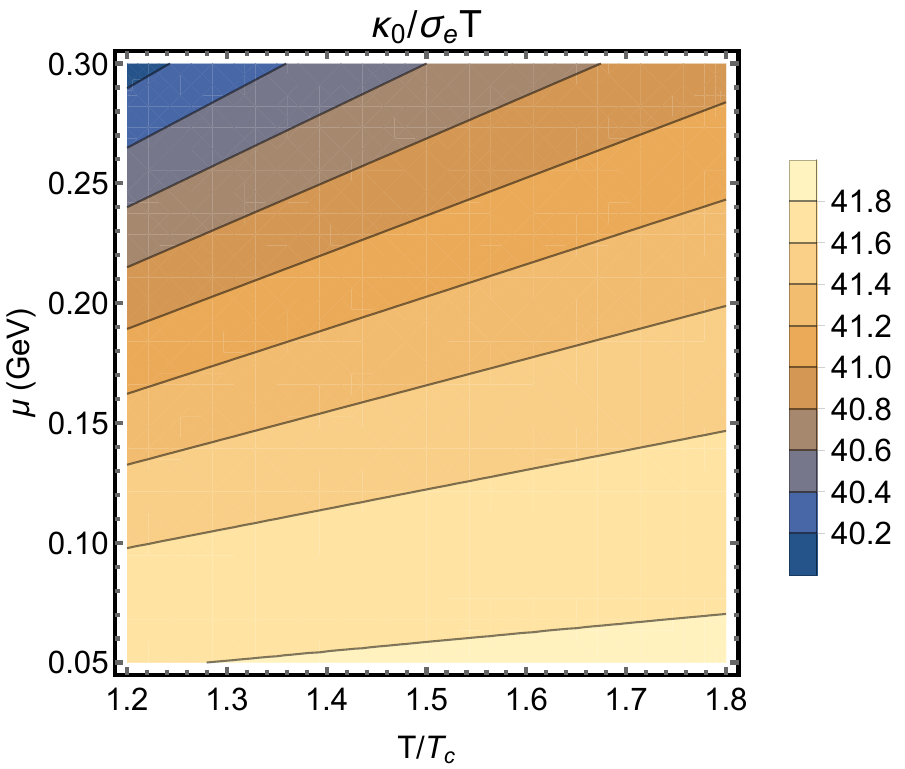}
    \hspace{-.2 cm}
    \includegraphics[width=0.43\textwidth]{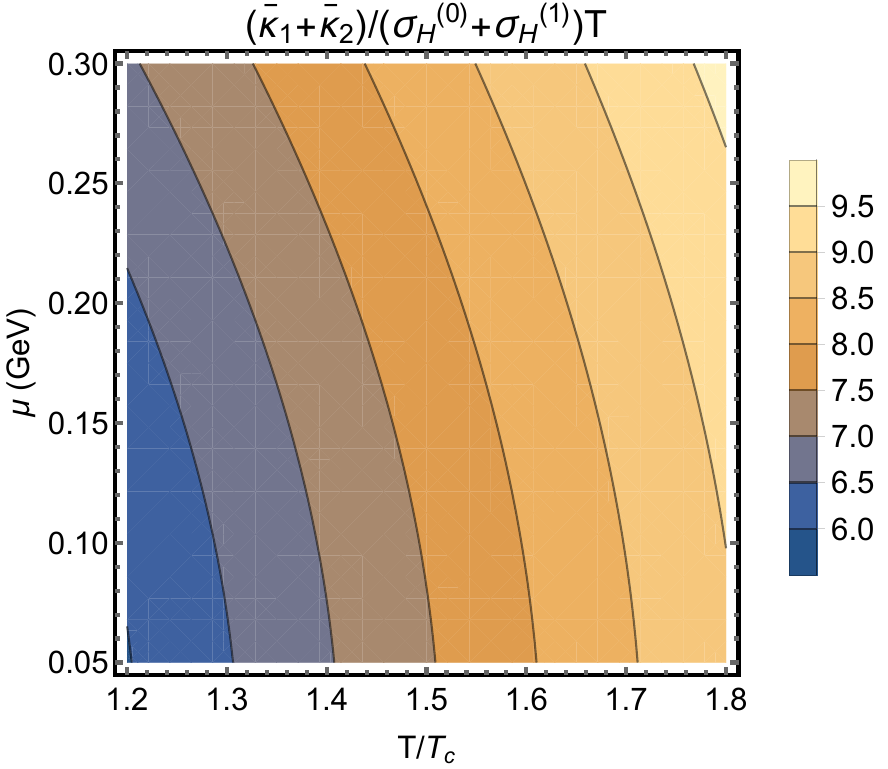}
    \caption{The relative significance of thermal to electric charge transport is plotted against temperature on the x-axis and chemical potential on the y-axis along two different directions, (left panel) $\frac{\kappa_0}{\sigma_e T}$  and (right panel) $\frac{\kappa_H}{\sigma_H T}$  for a constant value of magnetic field decay rate, $\tau_B=5$ fm. Curved lines denote constant value contours of depicted quantities.
 }
\label{f5}
\end{figure*}
\section{Impact of magnetic field on thermoelectric behavior of QCD medium}\label{IV}
In the presence of a time-varying magnetic field, there are different sources of the induced electric field in the conducting QGP medium. Most of the analyses consider the electric field due to the time decay of the magnetic field by using Faraday's law. Recently, it has been realized that the Seebeck effect can act as another source of the induced electric field, which is due to the local temperature gradient in the medium. Hence, the electric field induced in the medium ${\bf E}_{ind}$ can be expressed as,
{$${\bf E}_{ind} = {\bf E}_F+{\bf E}_T,$$
where ${\bf E}_F$ is the electric field due to Faraday's law, $\pmb{\nabla} \times \textbf{E} = \frac{\partial \textbf{B}}{\partial t}$ and ${\bf E}_T$ is the electric field induced due to the local temperature gradient in the medium}. 

We follow the same prescription as in Ref.~\cite{PhysRevD.102.014030} to explore the thermoelectric effect of the QGP medium. Here, we consider the case of an induced electric field from the temperature gradient, and hence we employ ${\bf E}_T\equiv {\bf E}$ with $E = |{\bf E}|$ in the rest of the analysis. The net current density of the QGP medium can be expressed as,
\begin{align}
{\bf j}&=2N_c\sum_f \int d P_k\,{\bf v}\,\Big(q_q f_q-q_{\bar{q}}f_{\bar{q}}\Big),  
 \end{align}
with the non-equilibrium part of the distribution function $\delta f_k= f_k-f^0_k$ takes the form as, 
\begin{align}\label{IV.3}
  \delta f_k = \textbf{p}.[\beta_1 \textbf{E} +\beta_2\textbf{B}+ \beta_3\textbf{X}+ \beta_4(\textbf{X}\times \textbf{B})+ \beta_5 \dot{\textbf{B}}\\ \nonumber +\beta_6(\textbf{X}\times \dot{\textbf{B}}) + \beta_7(\textbf{E}\times \textbf{B})+ \beta_8(\textbf{E}\times \dot{\textbf{B}})]\frac{\partial f^0_k}{\partial \epsilon_k}.  
\end{align}
The induced electric field can be obtained by considering the steady-state solution, $j =0$, to obtain a relation between $\textbf{E}$ and $\pmb{\nabla} T$. The electric field induced due to the temperature gradient in the presence of a magnetic field is along two directions,  say along the direction of $\pmb{\nabla} T$ and $\pmb{\nabla} T \times \textbf{B}$ respectively, and characterized by the Seebeck coefficient $S_B$ and Nernst coefficient $NB$, 
\begin{align}
 \begin{pmatrix}
E_x\\E_y
\end{pmatrix}
=
 \begin{pmatrix}
S_B & NB\\-NB &S_B
\end{pmatrix}
\begin{pmatrix}
 \frac{dT}{dx}\\ \frac{dT}{dy}
\end{pmatrix}.
\end{align}
The Seebeck and Nernst coefficients in the presence of a time-varying magnetic field can be defined as follows, 
\begin{align}
    &S_B = \frac{L_1 L_3 +L_2 L_4+L_2 L_6 +L_5 L_4 +L_5 L_6 }{L_1^2 +( L_2 + L_5)^2},\\
    &NB = \frac{L_1 L_4 + L_6 L_1 -L_2 L_3 -L_5 L_3}{L_1^2 +( L_2 + L_5)^2},
\end{align}
where the integrals $L_i, i=1,2..6$ take the following forms, 
\begin{align}
    &L_1 = \frac{2E}{3} N_c \sum_k \sum_f (q_{fk})^2 \int dP_k\frac{ p^2_k}{\epsilon_k^2}(-\frac{\partial f_k^0}{\partial \epsilon_k}) {N}_2,\\
    &L_2 = \frac{2E}{3} N_c \sum_k \sum_f (q_{fk})^3\int dP_k \frac{ p^2_k}{\epsilon_k^3}(-\frac{\partial f_k^0}{\partial \epsilon_k}) N_1,\\
    &L_3 = \frac{N_c}{3T} \sum_k \sum_f q_{fk} \int dP_k\frac{ p^2_k}{\epsilon_k^2} \frac{(\epsilon_k -h_k)M_1}{M_1^2 +M_2^2}(-\frac{\partial f_k^0}{\partial \epsilon_k}), \\
    &L_4 = \frac{N_c}{3T} \sum_k \sum_f q_{fk} \int dP_k\frac{ p^2_k}{\epsilon_k^2} \frac{(\epsilon_k -h_k)}{\tau_B (M_1^2 +M_2^2)}(-\frac{\partial f_k^0}{\partial \epsilon_k}),\\
    &L_5 = \frac{2E}{3\tau_B} N_c \sum_k \sum_f (q_{fk})^3\int dP_k \frac{ p^2_k}{\epsilon_k^3}(-\frac{\partial f_k^0}{\partial \epsilon_k}) \tau_R N_1,\\
    &L_6 = \frac{N_c}{3T} \sum_k \sum_f q_{fk} \int dP_k\frac{ p^2_k}{\epsilon_k^2} \frac{\tau_R (\epsilon_k -h_k)}{\tau_B^2(M_1^2 +M_2^2)}(-\frac{\partial f_k^0}{\partial \epsilon_k}).
\end{align}
We discuss the impact of the magnetic field, its time evolution, and chemical potential on the thermal and thermoelectric behavior of the QGP medium in the next section. 
\begin{figure*}
    \centering
    \includegraphics[width=0.45\textwidth]{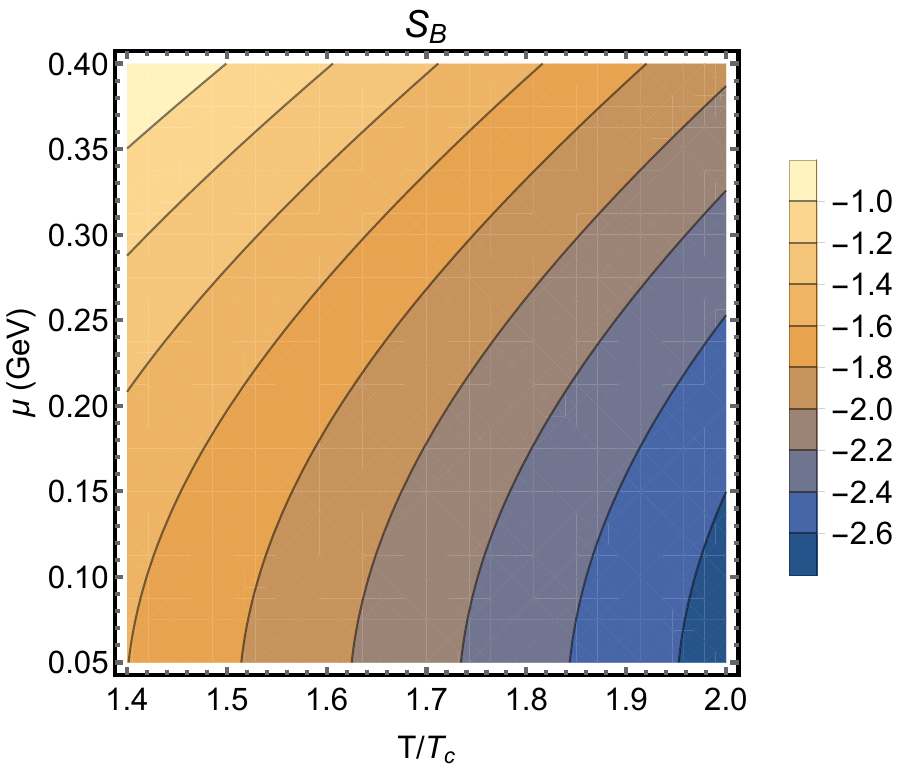}
    \hspace{-.2 cm}
    \includegraphics[width=0.44\textwidth]{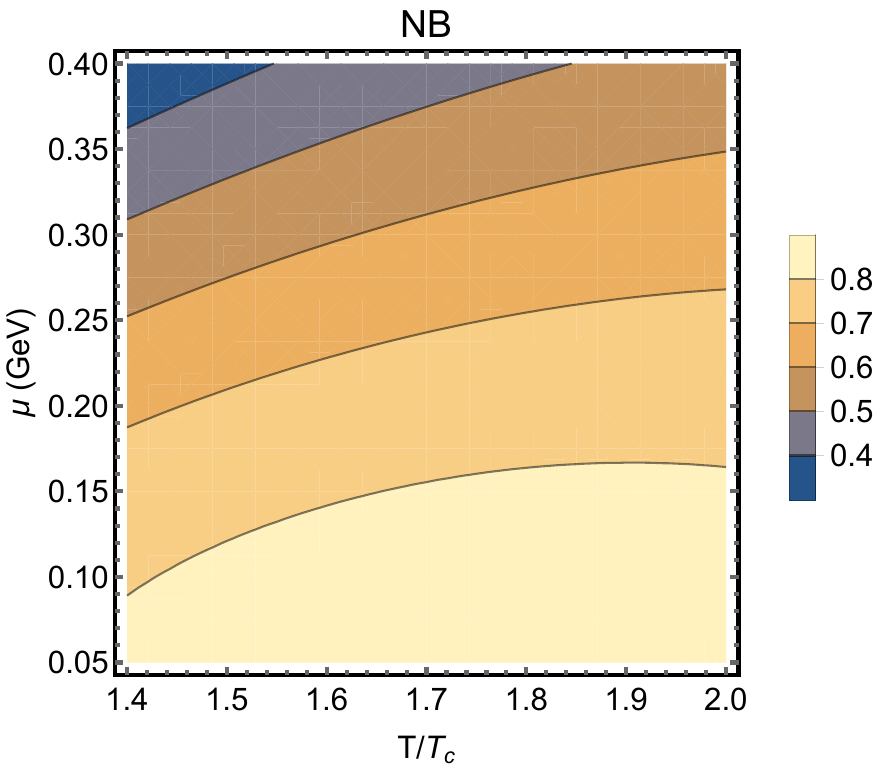}
    \caption{(Left panel) The Seebeck coefficient $S_B$ is plotted against temperature on the x-axis and chemical potential on the y-axis. (Right panel) The same plot for Nernst coefficient $NB$. The analysis has been carried out for a constant value of magnetic field decay rate, $\tau_B=5$ fm.}
\label{f6}
\end{figure*}

\section{Results and Discussions}\label{V}
We initiate the discussions with the transport coefficients associated with the thermal response of the QCD medium. The thermal transport process in the presence of a time-varying magnetic field is quantified by the thermal conductivity $\kappa_0$, which is leading order in terms of thermal relaxation time and `Hall-like' conductivities ($\bar{\kappa}_1$ and $\bar{\kappa}_2$) as described in Eq.~(\ref{1.111}). The impact of the time dependence of the magnetic field on the temperature behavior of the $\kappa_0$ with non-vanishing chemical potential is depicted in Fig.~\ref{f1} (left panel). The choice of the magnetic field evolution in the medium (with setup I and setup II) is shown to have a significant impact on thermal conductivity. The proper time evolution of $\kappa_0$ in the QCD medium for $eB_0=0.08$ GeV$^2$ at $\mu=100$ MeV is shown in Fig.~\ref{f1} (right panel).  Notably, the impact of the magnetic field evolution has more visible effects on the initial time.

In Fig.~\ref{f2} (left panel), the effect of magnetic field decay parameter $\tau_B$ and chemical potential on $\kappa_0$ is shown. Quantitatively, $\kappa_0$ increases with an increase in the decay parameter. This indicates that the magnetic field which decays slowly or persists longer in the QCD medium has more impact on the thermal transport of the medium. It is also important to emphasize that the thermal response of the medium is more visible in the high baryon density regimes. The magnetic field induces anisotropy in the thermal transport of the QCD medium and gives rise to heat current along the direction transverse to that of the magnetic field and temperature gradient. Relative significance of $\kappa_0$, $\bar{\kappa}_1$ and $\bar{\kappa}_2$ are shown in Fig.~\ref{f2} (right panel) for various values of $\tau_B$. The decay of the magnetic field is seen to have a significant role in the ratio, especially in the lower temperature regime.

The temperature behavior of the Knudsen number in the presence of a time-evolving magnetic field is depicted in Fig.~\ref{f3}. The impact of the time dependence of the magnetic field on the Knudsen number is seen to be more pronounced in the temperature regime near the transition temperature. The Knudsen number obtained preserves the continuum hypothesis, and hydrodynamics applies to the system of hot QCD matter as $Kn<<1$. It is seen that the Knudsen number increases with the decay rate of the magnetic field. A longer persisting magnetic field can affect the elliptic flow in the heavy-ion collision experiments. The dependence of elliptic flow $v_2$ on the temperature in a weakly magnetized medium is depicted in Fig.~\ref{f4}. The dependence of elliptic flow on the time dependence of the magnetic field is realized with various choices of $\tau_B$. We have observed that the longer the magnetic field persists, the larger its effect on $v_2$, and the effect is more pronounced in the low temperature regime.
We have compared the results with the observation from Ref.~\cite{Drescher:2007cd}.

The relative importance of thermal and electric charge transport in the QCD medium is analyzed using the Wiedemann-Franz law. The impact of chemical potential and temperature of the medium in the presence of a time-evolving magnetic field is shown in Fig.~\ref{f5}. As the magnetic field induces anisotropy in the medium, the Lorentz number is estimated in two different directions, say along the direction of the source of perturbations (thermal driving force and the electric field) and the direction perpendicular to the magnetic field in the medium. The thermal and the charge transport in the transverse direction of the magnetic field will have respective corrections due to the time dependence of the field in the medium as $\kappa_H = \bar{\kappa}_1 +\bar{\kappa}_2$ and $\sigma_H = \sigma_H^{(0)}+\sigma_H^{(1)}$. 
It is observed that the Wiedemann-Franz law is violated in both directions, and the value is dependent on the decay parameter of the magnetic field $\tau_B$.

In Fig.~\ref{f6}, the thermoelectric coefficients are plotted as a function of quark chemical potential and temperature in the presence of an evolving magnetic field. It is important to note that along with the temperature gradient, a non-vanishing chemical potential is required for the thermoelectric transport in the medium. The interplay of chemical potential and temperature on the Seebeck coefficient and Nernst coefficient is analyzed. It is observed that the temperature has a dominant role on the Seebeck coefficient than the chemical potential. However, for the case of the Nernst coefficient, the chemical potential seems to have a stronger dependence than the temperature of the medium. 
\section{Conclusion and Outlook}\label{VI}
We have explored the thermal and thermoelectric responses of the hot QCD medium in the presence of a time-varying magnetic field. We have obtained a general form of the heat current by solving the relativistic transport equation within the RTA for two different choices of magnetic field evolution in the medium. The thermal response due to the temperature gradient in the QCD medium has been quantified in terms of thermal conductivity. The decay time of the magnetic field and chemical potential in the medium seems to have a strong dependence on the induced heat current and temperature behavior of associated transport coefficients. The additional components of heat current due to the magnetic field and its time evolution in the medium have been explored in the analysis. The general framework of the thermal transport presented in the current study is consistent with other parallel studies and reproduce the results with constant and vanishing magnetic field with the appropriate choice of the field. 

The phenomenological relevance of the thermal transport of the medium with a time-evolving magnetic field has been investigated by studying elliptic flow. Further, we have analyzed the relative significance of thermal and electric charge transport in the hot QCD medium by evaluating the Lorenz number along two directions, along the direction of sources of perturbation and the direction perpendicular to that of the magnetic field. It is seen that the Wiedemann-Franz law is violated in the QCD medium, especially in the temperature regime near the transition temperature, in the presence of the time-dependent weak magnetic field. Finally, we have explored the dependence of magnetic field evolution on Seebeck and Nernst coefficients associated with the thermoelectric response of the QCD matter. The impacts of the decay time of magnetic field, chemical potential, and temperature on the thermoelectric coefficients of the QCD medium have been studied.

The thermal transport of the medium will be more relevant for upcoming experiments at FAIR as the baryon chemical potential will be significant in lower energetic collisions. The current analysis can be extended to explore the viscous coefficients of the QCD medium in the presence of inhomogeneous electromagnetic fields. The chirality, spin, and rotational effects of the medium on the electric charge transport is another interesting aspect to explore in the near future. 
\section*{Acknowledgments}
MK would like to acknowledge a scholarship from the Fonds de recherche du Quebec - Nature et technologies and support from Natural Sciences and Engineering Research Council of Canada.
\appendix
\section{\label{AppendixA} Calculation of $\alpha_i$}
In the present study, we consider the case of a slow-varying magnetic field to include the collisional aspects of the medium. Hence, the terms with  $\dot{\alpha_4}$ and $\dot{\alpha_5}$ are neglected in the present analysis as we are neglecting terms with second and higher order derivatives of the external perturbation. Comparing the tensorial structure on both sides of Eq.~(\ref{10}), we get,
\begin{align}\label{11.01}
  &\dot{\alpha_1} = -\frac{1}{\tau_R}\alpha_1+\frac{q_{f_k}(\textbf{B}.\textbf{X})}{\epsilon_k}\alpha_3\\
  & \dot{\alpha_2} =  -\bigg[\frac{1}{\tau_R}\alpha_2 +(\frac{q_{fk}(\textbf{B}.\textbf{B}-\textbf{B}.\dot{\textbf{B}})}{\epsilon_k})\alpha_3 -\frac{\epsilon_k -h_k}{\epsilon_k}\bigg], \\  
  & \dot{\alpha_3} = -\frac{1}{\tau_R}\alpha_3 +\frac{q_{fk}}{\epsilon_k}\alpha_2,\label{11.1}
\end{align}
along with the coupled equations,
\begin{align}
&{\alpha_4}=-{\tau_R}\Big[\alpha_1+\frac{\tau_R q_{f_k} (\textbf{B}.\textbf{X})}{\epsilon_k} \alpha_5 \Big], &&\alpha_5 = -\tau_R \alpha_3.
\end{align}
Further, Eq.~(\ref{11.01})-(\ref{11.1}) can be expressed in terms of matrix equation as,
\begin{equation}\label{12.01}
 \frac{d X}{d t}= AX +G,   
\end{equation}
where the matrices take the following forms,
\begin{align}
 X=\begin{pmatrix}
\alpha_1\\\alpha_2\\\alpha_3
\end{pmatrix},  
&&A=\begin{pmatrix}
-\frac{1}{\tau_R} &0 &\frac{q_{f_k}}{\epsilon_k}(\textbf{B}.\textbf{X})\\
 0 &-\frac{1}{\tau_R} & -\frac{q_{f_k} F^2}{\epsilon_k}\\
0 &\frac{q_{f_k}}{\epsilon_k} &  -\frac{1}{\tau_R}, 
\end{pmatrix},
\end{align}
\begin{align}
&G=\begin{pmatrix}
0\\ \frac{\epsilon_k -h_k}{\epsilon_k}\\ 0
\end{pmatrix},\nonumber
\end{align}
where $F = \sqrt{B(B-\tau_R \dot{B})}$. Eq.~(\ref{12.01}) can be solved by diagonalizing the matrix $A$ and employing the variation of constants method by considering the parameters $c_1$, $c_2$ and $c_3$ to be dependent on time as, $c_1(\tau)$, $c_2(\tau)$ and $c_3(\tau)$. The solution is as follows,  
\begin{align}\label{14}
&\alpha_1 = c_1 e^{\eta_1} +\zeta c_2 e^{\eta_2} -\zeta c_3 e^{\eta_3},
\end{align}
\begin{align}
&\alpha_2 =-c_2 iFe^{\eta_2} +c_3 iFe^{\eta_3}, &&\alpha_3=c_2 e^{\eta_2} +c_3 e^{\eta_3}.\nonumber
\end{align}
The functions $c_1(\tau)$, $c_2(\tau)$ and $c_3(\tau)$ can be defined as $c_1 =-i\frac{(\epsilon_k -h_k)}{\epsilon_k} \zeta I_1$,  $c_2 =i\frac{(\epsilon_k -h_k)}{2\epsilon_k} I_2$ and $c_3 =-i\frac{(\epsilon_k -h_k)}{2\epsilon_k} I_3$  for the parameter $\zeta = \frac{i(\textbf{B}.\textbf{X})}{F}$. The integrals $\eta_j$ and $I_j$ $(j=1,2,3)$, take the following forms, 
\begin{align}\label{15}
    &\eta_j = -\frac{\tau}{\tau_R} +a_j\frac{q_{f_k} i}{\epsilon_k}\int F d\tau,
    &&I_j = \int \frac{ e^{-\eta_j}}{F}(\omega -h_k)d\tau,
\end{align}
with $a_1=0$, $a_2=1$ and $a_3 = -1$. Substituting Eq.~(\ref{15}) in Eq.~(\ref{14}), we obtain the form of $\alpha_i$.
\section{\label{AppendixB} Calculation of $\beta_i$}
Substituting the form of $\delta f_k$ as defined in Eq.~(\ref{IV.3}) in the Boltzmann equation with the RTA collision kernel, 
\begin{align}
\frac{\partial f_k}{\partial t} +{\bf v}.\frac{\partial f_k}{\partial {\bf x}} +q_{f_k} (\textbf{E}+ \textbf{v} \times \textbf{B}).\frac{\partial f_k}{\partial {\bf p}} =-\frac{\delta f_k}{\tau_R}.
\end{align}
Following the same formalism as in appendix \ref{AppendixA}, the coefficients $\beta_i$, $i=1...8$ can be described as follows,
\begin{align}\label{}
 &\beta_1  =-\frac{\Tilde{\Omega}_k}{2}(I_1 e^{\eta_1} +I_2 e^{\eta_2} ),\\
 &\beta_2 =-\frac{i(\textbf{B}.\textbf{X})}{F \epsilon_k} I_0 e^{\eta_0}+\frac{i(\textbf{B}.\textbf{X})}{2F \epsilon_k} I_1 e^{\eta_1} +\frac{i(\textbf{B}.\textbf{X})}{2F \epsilon_k} I_2 e^{\eta_2},\\ 
 &\beta_3 =-i\frac{F}{2 \epsilon_k} I_1 e^{\eta_1}+i\frac{F}{2 \epsilon_k} I_2 e^{\eta_1} ,\label{20.2}\\
 &\beta_4 =\frac{ I_1}{2 \epsilon_k} e^{\eta_1} +\frac{I_2}{2 \epsilon_k} e^{\eta_1},\label{20.4}\\
 &\beta_5 = -\tau_R \alpha_1 -\frac{q_{fk} \tau_R^2}{\epsilon_k} \alpha_3 (\textbf{B}.\textbf{X}),\label{20.1}\\
 &\beta_6 =-\tau_R \alpha_3,\\
 &\beta_7 = \frac{q_{fk}i}{2\epsilon_k}(I_1 e^{\eta_1} -I_2 e^{\eta_2} ),\\
 & \beta_8 = -\frac{\tau_R q_{fk}i}{2\epsilon_k}(I_1 e^{\eta_1} -I_2 e^{\eta_2}  ),
\end{align}
with, $\Tilde{\Omega}_k=\frac{q_{f\,k}F}{\epsilon}$.

\bibliography{ref}{}

\end{document}